\documentclass[aps,prd,preprint,superscriptaddress,tightenlines,nofootinbib]{revtex4}

\usepackage{graphicx}
\usepackage{dcolumn}
\usepackage{bm}
\usepackage{epsfig}
\usepackage[dvips]{color}
\usepackage{graphics,color,epsfig}

\newcommand{\bodyspacing}{1}
\newcommand{\bbsp}{\renewcommand{\baselinestretch}{\bodyspacing}\small\normalsize}
\newcommand{\ebsp}{\par\renewcommand{\baselinestretch}{1}\small\normalsize}

\newcommand{\bfigenvh}{\ebsp\begin{figure}[h]}
\newcommand{\bfigenvhh}{\ebsp\begin{figure}[h!]}
\newcommand{\bfigenv}{\ebsp\begin{figure}}
\newcommand{\efigenv}{\end{figure}\bbsp}
\newcommand{\btabenvh}{\ebsp\begin{table}[h]}
\newcommand{\btabenv}{\ebsp\begin{table}}
\newcommand{\etabenv}{\end{table}\bbsp}
\newcommand{\bi}{\ebsp\begin{itemize}}
\newcommand{\ei}{\end{itemize}\bbsp}

\newcommand{\bc}{\begin{center}}
\newcommand{\ec}{\end{center}}
\newcommand{\be}{\begin{equation}}
\newcommand{\ee}{\end{equation}}

\def\mtiny{\vrule width 0pt}
\def\mrm#1{\mathrm{#1}}
\def\DZ{\relax\ifmmode{D^0}\else{$\mrm{D}^{\mrm{0}}$}\fi}
\def\Dz{\relax\ifmmode{D^0}\else{$\mrm{D}^{\mrm{0}}$}\fi}
\def\DZB{\relax\ifmmode{\overline{D}\mtiny^0}
        \else{$\overline{\mrm{D}}\mtiny^{\mrm{0}}$}\fi}
\def\Dzb{\relax\ifmmode{\overline{D}\mtiny^0}
        \else{$\overline{\mrm{D}}\mtiny^{\mrm{0}}$}\fi}
\def\KZ{\relax\ifmmode{K^0}\else{$\mrm{K}^{\mrm{0}}$}\fi}
\def\KZB{\relax\ifmmode{\overline{K}\mtiny^0}
        \else{$\overline{\mrm{K}}\mtiny^{\mrm{0}}$}\fi}
\def\BZ{\relax\ifmmode{B^0}\else{$\mrm{B}^{\mrm{0}}$}\fi}
\def\BZB{\relax\ifmmode{\overline{B}\mtiny^0}
        \else{$\overline{\mrm{B}}\mtiny^{\mrm{0}}$}\fi}
\def\DZS{\relax\ifmmode{D^{*+}}\else{$\mrm{D}^{\mrm{*+}}$}\fi}
\def\DZM{\relax\ifmmode{D^{*-}}\else{$\mrm{D}^{\mrm{*-}}$}\fi}
\def\DZC{\relax\ifmmode{\overline{D}\mtiny^0}
        \else{$\overline{\mrm{D}}\mtiny^{\mrm{0}}$}\fi}
\def\DZD{\relax\ifmmode{{\bar D}\mtiny^0}
        \else{${\bar{\mrm{D}}}\mtiny^{\mrm{0}}$}\fi}

\begin{document}

{\def\mtiny{\vrule width 0pt}
\def\DZ{\relax\ifmmode{D^0}\else{$\mrm{D}^{\mrm{0}}$}\fi}
\def\DZB{\relax\ifmmode{\overline{D}\mtiny^0}\else{$\overline{\mrm{D}}\mtiny^{\mrm{0}}$}\fi}
}

\preprint{CLNS 05/1916}
\preprint{CLEO 05-8}         

\title{ Searches for $CP$ Violation and $\pi\pi$ $S$-wave in the Dalitz-Plot
Analysis of  $\DZ
\rightarrow \pi^+ \pi^- \pi^0$}

\author{D.~Cronin-Hennessy}
\author{K.~Y.~Gao}
\author{D.~T.~Gong}
\author{J.~Hietala}
\author{Y.~Kubota}
\author{T.~Klein}
\author{B.~W.~Lang}
\author{S.~Z.~Li}
\author{R.~Poling}
\author{A.~W.~Scott}
\author{A.~Smith}
\affiliation{University of Minnesota, Minneapolis, Minnesota 55455}
% ...
\collaboration{CLEO Collaboration} 
\noaffiliation


\date{March 28, 2005}

\begin{abstract} 
In $e^+e^-$ collisions recorded using the CLEO~II.V detector we have studied the
Cabibbo suppressed decay of $\DZ \to  \pi^+\pi^-\pi^0$ with the initial
flavor of the $\DZ$ tagged by the decay $D^{*+}\to \DZ\pi^+$.
We use the Dalitz-plot analysis technique to measure the resonant 
substructure in this final state and observe $\rho\pi$ and non-resonant
contributions by fitting for their amplitudes and relative phases.
We describe the $\pi\pi$ $S$-wave with a $K$-matrix
formalism and limit this contribution to the rate to be $< 2.5\%$ @
95\% confidence level,
in contrast to the large rate observed in $D^+\to \pi^+\pi^-\pi^+$ decay.
Using the amplitudes and phases from this analysis, we calculate an
integrated $CP$ asymmetry of $0.01^{+0.09}_{-0.07} \pm 0.05 $.  
\end{abstract}

\pacs{13.25.Ft}
\maketitle

The light scalar meson sector is an enduring puzzle in QCD~\cite{pdgscalar}.
Specifically the isospin zero, $J^{PC}=0^{++}$ mesons are complex
both theoretically and experimentally.  The singly-Cabibbo suppressed
decays of $D$ mesons are an excellent laboratory to test this sector.
The final state consists of only $u$ and $d$ quarks
and antiquarks so there is sufficient energy to cover most of the
range of interest to light quark binding, and the initial state is 
simple with little impact on the final state. 
A better understanding of final state interactions in
exclusive weak decays is needed in order to model rates,
explain interesting phenomena such as mixing~\cite{ref:hepph9802291}
and elucidate the origin of $CP$ violation in the $B$ sector~\cite{Giri:2003ty}.

Weak decays of $D$ mesons are expected
to be dominated by resonant two body decays~\cite{ref:bsw,ref:bedaque,ref:chau,ref:terasaki,Buccella}. The well established
Dalitz-plot analysis technique can be used to explore the resonant 
substructure which should be rich in isospin zero mesons.  Recently the FOCUS
collaboration studied the Dalitz plot $D^+ \to
\pi^+\pi^-\pi^+$~\cite{focus3pi} and
observed a large $\pi\pi$ $S$-wave contribution using a $K$-matrix 
formulation to describe to $0^{++}$ resonance
structures. This letter describes a similar Dalitz-plot analysis of 
$D^0 \to \pi^+\pi^-\pi^0$ at CLEO in which $\pi\pi$ $S$-wave
contributions are also expected. We have searched for such 
contributions using the $K$-matrix formulation following what has
been done by FOCUS and alternatively for resonance contributions from the
scalar $\sigma(500)$ and $f_0(980)$ mesons.  We see no evidence for
any $\pi\pi$ $S$-wave contribution in $D^0 \to \pi^+\pi^-\pi^0$, and 
fully describe the Dalitz plot with contributions from $\rho$ resonances.

Standard Model (SM) predictions for the rate of $CP$ violation in 
almost all charm meson decay modes is ${\cal O} (10^{-6})$.
However, for some singly Cabibbo suppressed decays of $D$ mesons such
as $\DZ \to \pi^+\pi^-\pi^0$, the
SM predictions for the rate of $CP$ violation are as large as
0.1\%~\cite{Buccella,Santorelli}, due to interference
between tree and penguin processes.

Previous investigations~\cite{MarkIII} of this decay 
were limited by statistics and did not search for $CP$ violation nor
study the resonant substructure.

This analysis uses an integrated luminosity of 9.0~fb$^{-1}$
of $e^+e^-$ collisions at $\sqrt{s}\approx10\,$GeV provided by
the Cornell Electron Storage Ring (CESR).
The data were taken with the CLEO~II.V 
detector~\cite{cleoii,cleoiiv}. 

We reconstruct candidates for the decay sequence
$D^{\ast+}\!\to\!\pi^+_s \DZ$, $\DZ\!\to\pi^+\pi^-\pi^0$.
Charge conjugation
is implied throughout this Letter.
The charge of the slow
pion ($\pi^+_s$ or $\pi^-_s$) identifies the charm state 
at $t=0$ as either $\DZ$ or $\DZB$. 
To reduce background, we require the $D^{\ast+}$ momentum
to exceed 70\% of its maximum value $\sqrt{E^2_{beam}-M^2_{D^{*+}}}$. 
The $\pi^0$ candidates are reconstructed from all pairs of
electromagnetic showers that are not associated with charged tracks.
To reduce the number of fake $\pi^0$ from random shower combinations 
we require that each shower has energy greater than 100~MeV and be in
the barrel region of our detector. The two photon invariant mass
is required to be $120< M_{\gamma\gamma} < 150$~MeV$/c^2$. 
To improve the mass resolution, the invariant mass is constrained to
the known
$\pi^0$ mass and we require the $\chi^2$ of this fit to be $<100$.
We exploit the precision tracking of the silicon vertex detector~\cite{cleoiiv} by 
refitting the 
$\pi^\pm$ tracks with a requirement that they
form a common vertex in three dimensions.
We use the trajectory of the
$\pi^+\pi^-\pi^0$ system and the position of the CESR luminous region
to obtain the $\DZ$ production point.  
We then refit the $\pi_s^+$ track with a requirement that
the trajectory intersect
the $\DZ$ production point.

\begin{figure}[t]
\begin{center}
\epsfig{figure=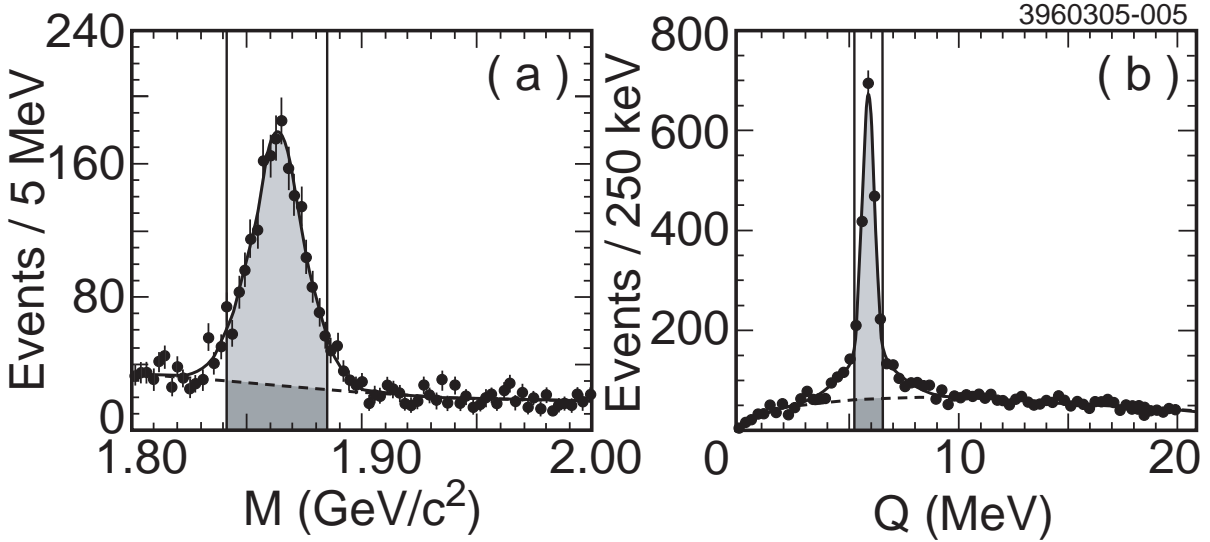,width=.99\textwidth}
\end{center}
\caption{Distribution of a)
$M$ within 650~keV of our measured value of $Q$ and b) $Q$
within $44$~MeV/$c^2$ of our measured value of $M$
for the process $\DZ \to \pi^+\pi^-\pi^0$.
The candidates  
pass all selection criteria discussed in the text.
The curves show the results of the fits. 
The vertical lines denote the signal region. The light (dark)
shaded region indicates the signal (background) contribution.
}
\label{fig:qandm}
\end{figure}

We reconstruct the energy released in the 
$D^{\ast+}\!\to\!\pi^+_s\DZ$ decay as $Q\!\equiv\!M^\ast\!-\!M\!-\!m_\pi$, 
where $M^\ast$ is the reconstructed mass of the
$\pi_s^+ \pi^+ \pi^-\pi^0$ system, $M$ is the reconstructed mass of the
$\pi^+\pi^-\pi^0$ system, and $m_\pi$ is the charged pion mass.  
The addition of the $\DZ$ production point to the $\pi^+_s$ trajectory
improves the resolution on $Q$ by a factor of two.
The distributions of $Q$ and $M$ for our data are shown in
Fig.~\ref{fig:qandm}. 
We fit the $M$ and $Q$ distributions separately to a double Gaussian
plus a
background shape and find an average background fraction of $(18.6 \pm
3.6)$\%.  
We select 1917 candidates within 650~keV of the nominal 
value of $Q$, denoted as $\bar Q$, and within 44~MeV/$c^2$ of the
nominal value of $M$, both as measured in this analysis.

The efficiency for the selection
described above is not uniform across the Dalitz-plot distribution
($m_{\pi^+\pi^-}^2$, $m_{\pi^+\pi^0}^2$).
We study the efficiency with a GEANT~\cite{GEANT} based simulation of
the detector with a luminosity corresponding
to more than twenty times our data sample.
To measure the variation in efficiency over the Dalitz plot,
we generate signal Monte Carlo 
uniformly populating the allowed phase space.  
We observe deviations from the uniform distribution due to momentum
dependent $\pi^0$ reconstruction efficiency and 
inefficiencies near the edge of phase space. The average reconstruction
efficiency is $\sim$3.7\% but increases (decreases) to $\sim$4.5\%
($\sim$0.5\%) for decays with high (low) momentum $\pi^0$ mesons. 
We fit the efficiency to a two dimensional cubic
polynomial in ($m_{\pi^+\pi^-}^2$, $m_{\pi^+\pi^0}^2$).

	Figure~\ref{fig:qandm} shows that the background is significant.
To construct a model of the background shape, we consider 
events in the data in sidebands $3 < \bar Q - Q < 6$~MeV and
$3 < Q - \bar Q < 15$~MeV
within the $M$ signal region defined above.
There are 2711 events in this selection, about eight times the amount
of background we estimate from the signal region. The background is
dominated by random combinations of unrelated tracks and
showers. Although the background includes $\rho$ 
and $K^0_S$ mesons combined with random tracks and/or showers,
these events will not interfere with each other or
with resonances in the signal as they are not from a true $\DZ$. 
Additionally, $K^0_S \pi^0$ events populate a narrow region on the
Dalitz plot in both
signal and background.  The corresponding amplitudes do not
interfere with the other amplitudes that contribute to 
$\DZ \to \pi^+\pi^-\pi^0$ due to the long lifetime of the $K^0_S$.
Therefore, the normalization of the $K^0_S$
contribution component floats in the fit, but has no further role.
The background shape is parameterized by a two dimensional cubic
polynomial in ($m_{\pi^+\pi^-}^2$, $m_{\pi^+\pi^0}^2$) with terms
representing $\rho$ and $K^0_S$ mesons.

\begin{figure}[t]
\epsfig{figure=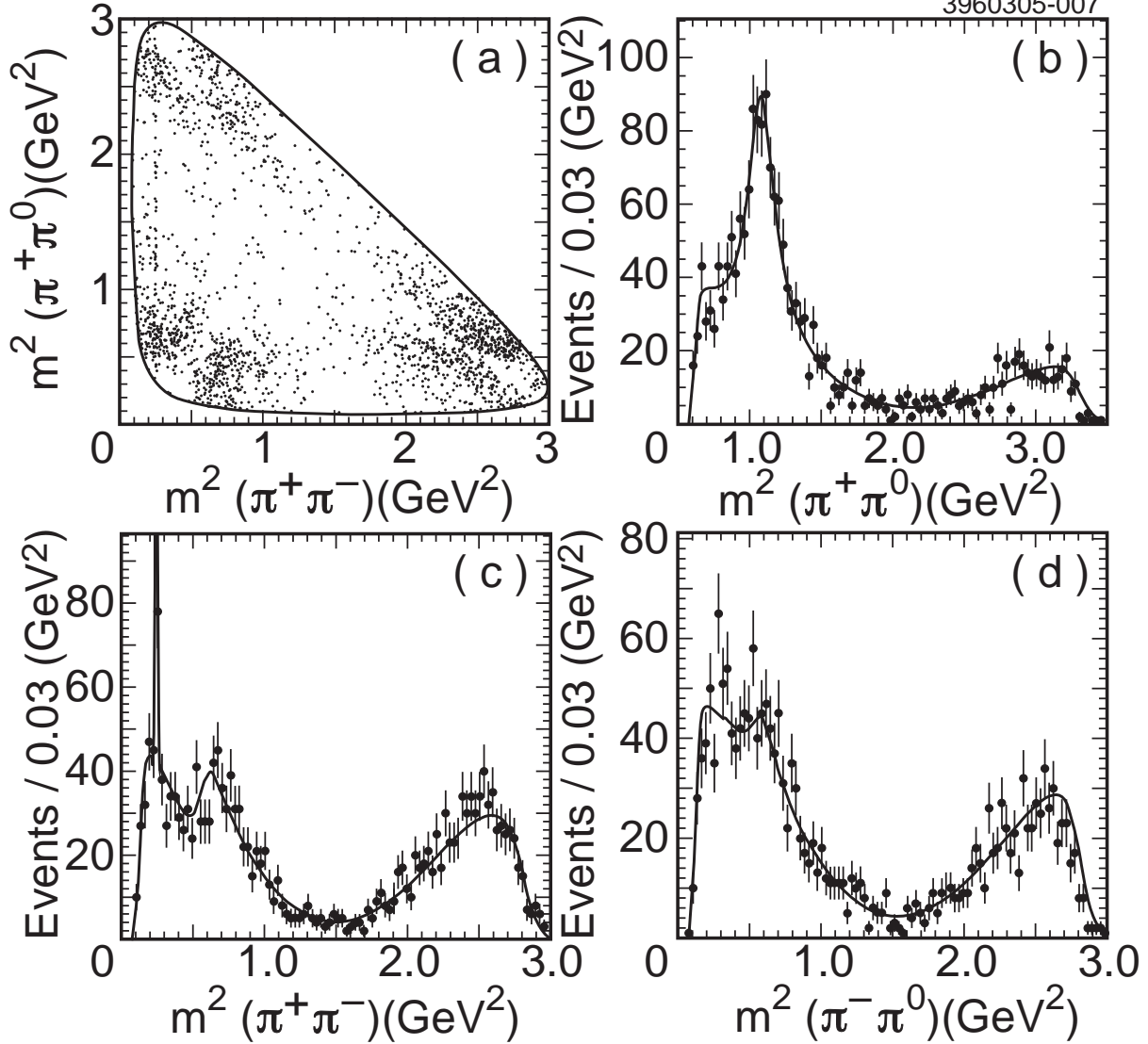,width=.99\textwidth} 
\caption
{\label{fig:proj}
a) The Dalitz-plot distribution for $\DZ\to \pi^+\pi^-\pi^0$ 
candidates. b)-d) Projections of the results of the fit described in the text
to the $\pi^+\pi^-\pi^0$~Dalitz plot showing both 
Fit~D (line) and the data (points). The results of Fit~A, Fit~C1,
and Fit~C2 are indistinguishable from Fit~D. See text for details of
the fits.}
\end{figure}

Figure~\ref{fig:proj} shows the Dalitz-plot distribution for the
$\DZ \to \pi^+\pi^-\pi^0$ candidates.  
Only contributions from $\rho^\mp\pi^\pm$, $\rho^0\pi^0$ and 
$K^0_S\pi^0$ are readily apparent.
Modeling the background shape
and correcting for efficiency as described above, 
we then parameterize the $\pi^+\pi^-\pi^0$ Dalitz-plot distribution following 
the Breit-Wigner formalism using the unbinned likelihood method as
described in Ref.~\cite{tjb,cleokspipi}. In a separate 
fit, we also parameterize the $\pi\pi$ S-wave with the $K$-matrix
formalism as described in Ref.~\cite{focus3pi} for the analysis of
$D^+\to \pi^+\pi^-\pi^+$. We allow the normalization of the
background contribution to float unconstrained in our fits.
We consider seventeen resonant components, 
$\sigma\pi^0$, 
$\rho^0 \pi^0$,
$\rho^\mp \pi^\pm$,
$\omega \pi^0$,
$f_0(980) \pi^0$,
$f_2(1270) \pi^0$,
$f_0(1370) \pi^0$,
$\rho(1450)^0 \pi^0$,
$\rho(1450)^\mp \pi^\pm$,
$f_0(1500) \pi^0$,
$\rho_3(1690)^0 \pi^0$,
$\rho(1700)^0 \pi^0$,
$\rho(1700)^\mp \pi^\pm$,
$f_0(1710) \pi^0$,
as well as a non-resonant contribution.  All interfere coherently
and we fit for a complex coefficient (amplitude and relative phase)
for each resonance as well as for the non-resonant contribution.  
We describe the resonances with the standard parameters~\cite{pdg}.
Lacking theoretical guidance, the non-resonant contribution is 
modeled as a uniform distribution across the allowed phase space.  

This study is sensitive only to relative phases 
and amplitudes.
The largest mode, $\rho^+ \pi^-$, is assigned
a zero phase and an amplitude of one.  
Since the choice of normalization, phase convention, and amplitude
formalism may not always be identical for different experiments, 
fit fractions are reported in addition to amplitudes.
The fit fraction is defined as the integral of a single
component (resonant or non-resonant) over the Dalitz plot divided by the integral of the coherent
sum of all components over the Dalitz plot~\cite{tjb}.
The sum of the fit fractions for all components
will in general not be unity because of interference.

We use the full covariance matrix from the fits to determine
the statistical errors on fit fractions to properly
include the correlated components of the uncertainty on the amplitudes and
phases. After each fit, the covariance matrix and final parameter
values are used to generate sample parameter sets.  
The distributions of fit fractions from these parameter sets
are then used to determine the Gaussian width and 95\% confidence
level (C.L.) upper limits.

The results of our fits are presented in Table~\ref{tab:fit}. 
Fit~A includes the three $\rho(770)$ resonances
and an interfering non-resonant component.
The non-resonant contribution is small and we do not find strong
evidence for any other contributing resonances. 

We fit the $\DZ$ and $\DZB$ samples separately in Fit~B1 and Fit~B2, 
respectively. 
The violation of $CP$ could manifest as distinct amplitudes and phases
for $\DZ$ and $\DZB$ Dalitz plots. Since Fits B1 and B2 are consistent,
there is no indication of $CP$ violation.  

\begin{figure}[t]
\epsfig{figure=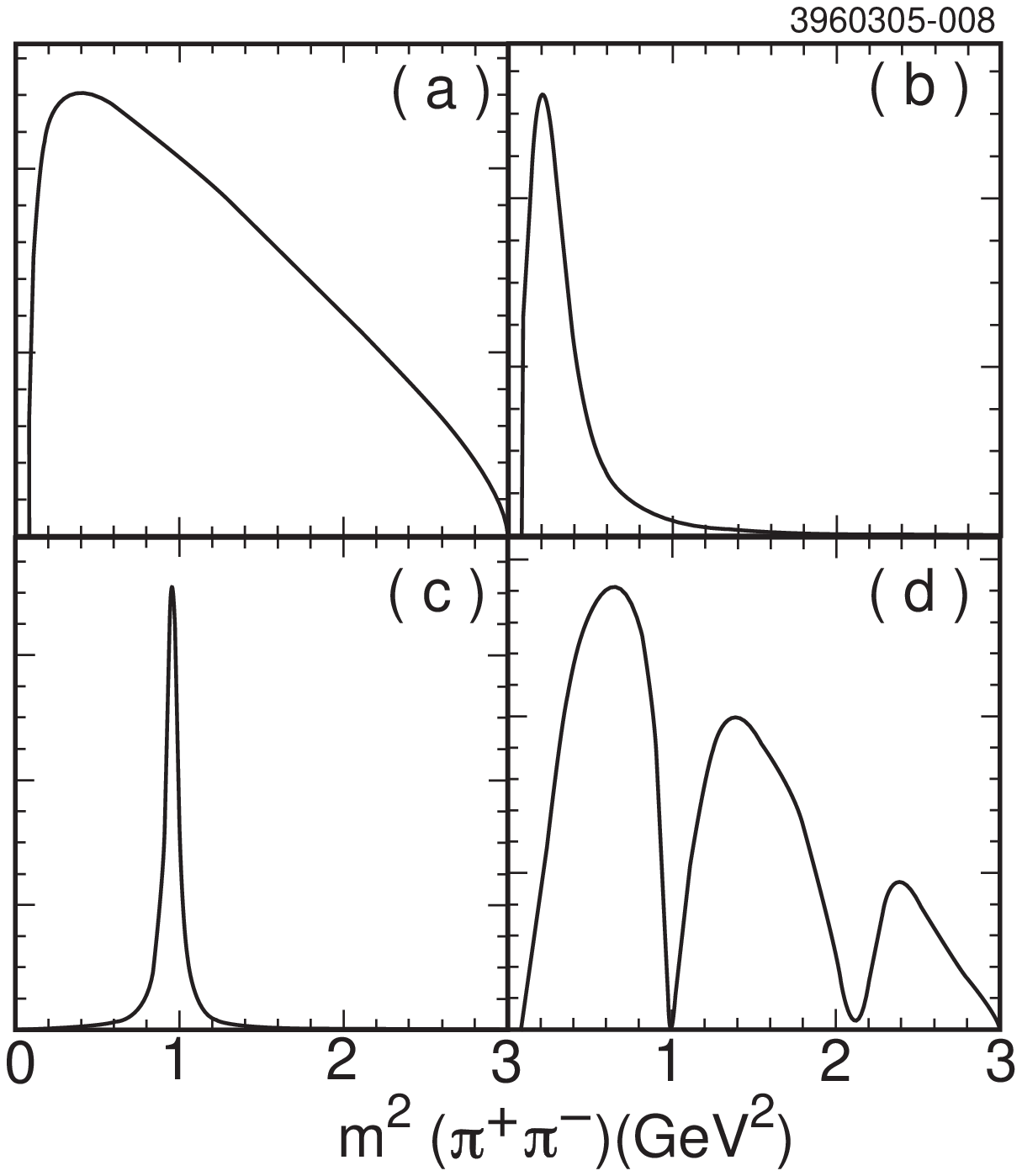,width=.99\textwidth}
\caption
{\label{fig:Swave}
Projections of the $\pi\pi$ $S$-wave parameterization for a) Fit~A:
flat non-resonant b) Fit~C1: Breit-Wigner $\sigma(500)$ c) Fit~C2:
Breit-Wigner $f_0(980)$ and d) Fit~D: K-matrix.}
\end{figure}

There has been significant interest in the properties
of the $\pi\pi$ $S$-wave due to the possibility of a low mass $\sigma$
meson and glueball of mass $\sim$ 1.5 GeV/$c^2$, each of which are 
beyond the $q\bar q^\prime$ quark model~\cite{qqbarmodel}.  
A recent paper~\cite{Oller} highlights $D$ decay as a preferred way
to focus on these states because of the preponderance of $S$-waves
in the initial and final state.  Charged $D$
mesons are observed to decay
preferentially to the $\pi\pi$ $S$-wave in the decay $D^+\to \pi^+\pi^-\pi^+$
studied by FOCUS~\cite{focus3pi} and E791~\cite{e791}.  
We considered several possible contributions of a $\pi\pi$ 
$S$-wave component shown in Fig.~\ref{fig:Swave}. 
In Fit~C1 we replace the flat non-resonant amplitude with a 
$\sigma(500)$~\cite{pdg}
contribution parameterized as a Breit-Wigner resonance.
The contribution of $\sigma$ is very small in our fit.

In Fit~C2, we replace the $\sigma(500)$ with the $f_0(980)$ where we
use the mass and width determined from the single Breit-Wigner
parameterization of E791~\cite{e791f0}.  
There are predictions~\cite{Buccella} for 
$D^+ \to f_0(980)\pi^+$ (BF=2.8\%) and $\DZ \to f_0(980) \pi^0$ (BF=0.06\%).
The prediction for the latter decay is extremely small because
of the non-$q\bar q$ nature of $f_0(980)$ and strong 
final state interactions. 
We find a fit fraction of $(0.010 \pm 0.008)$\% for $\DZ \to f_0(980) \pi^0$.
Although statistical errors limit the precision of our result,
it is clear that this decay is highly suppressed in agreement with the model.

The $\pi\pi$ $S$-wave has contributions
from a number of overlapping resonances and there are several
models which parameterize this wave from other 
reactions~\cite{aumope,sarantsev,focus3pi}.  The complicated
structure is much more amenable to a coupled channel formulation
than Breit-Wigner models.  In Fit~D
we consider a $\pi\pi$ $S$-wave contribution following the $K$-matrix 
formalism of Au, Morgan, and Pennington~\cite{aumope} in addition 
to the resonant components of Fit~A.  
As in Fit~C1 and Fit~C2, we exclude the non-resonant amplitude.  
Figure~\ref{fig:proj}b-d shows the three projections of the Fit~D.
These fits are very similar to those of FOCUS~\cite{focus3pi}
for $D^+\to \pi^+\pi^-\pi^+$. 
The fit fraction of the $\pi\pi$ $S$-wave component in Fit~C1 is much
less than that of E791 for the analogous fit to $D^+\to
\pi^+\pi^-\pi^+$.  
For Fit~D, our fit
fraction, $(0.9 \pm 0.7)$\%, was very small  
when compared to the FOCUS value
for $D^+$ decay, $(56.00 \pm 3.24 \pm 2.08)$\%.  
The observed ratio of $\pi\pi$ $S$-wave fit fractions in $D^+$
relative to $\DZ$ is $60^{+51}_{-17}$ or $>\!36\,@ 95\%$ C.L. 
- somewhat larger than the tree-level estimate
of $(3\sqrt{2})^2=18$. The factors in the relative amplitudes of 3 and $\sqrt{2}$ are due to color
suppression and isospin, respectively.

The lack of evidence for a $\pi\pi$ $S$-wave in this analysis is 
interesting.  The measured rates for the comparable reactions
$D^+ \to \pi^+\pi^0$, $(0.26 \pm 0.07)$\%, and $\DZ \to \pi^0 \pi^0$,
$(0.084 \pm 0.022)$\%, are both given qualitatively by model 
calculations~\cite{Buccella} which provide values of 0.19\% and
0.11\%.  A quark model with final state interactions is used to 
fit parameters to a few charm decays and then predict many others.  
The same model predicts a ratio
of branching fractions (BF)
for $D^+ \to f_0(980)\pi^+$ and $\DZ \to f_0(980) \pi^0$
of 46.7.  Comparing the results from Ref.~\cite{e791f0} and our
Fit~C2 we measure this ratio to be $620^{+620}_{-210}$ or $>\!340$ @ 95\% C.L..
Although the quantities measured in the experiments are 
not the same as what is calculated in the model, there is a strong 
likelihood that $D$ three-body decay data can provide
interesting tests of the microscopic nature of the scalar states.

\begin{table}
\caption{Fit~A includes the three $\rho(770)$ resonances
and a non-resonant (NR) contribution.
We fit the $\DZ$ and $\DZB$ samples separately in Fit~B1 and Fit~B2, 
respectively.
Fit~C1 and Fit~C2 allow a $\sigma(500)$ and $f_0(980)$
contribution, respectively, parameterized as a Breit-Wigner resonance~\cite{tjb}.
The $\pi\pi$ $S$-wave contribution in Fit~D is
parameterized following the $K$-matrix formalism~\cite{focus3pi}.}
\label{tab:fit}
\begin{tabular}{cccc} \hline \hline
 & Amplitude & Phase($^\circ$) & Fit fraction(\%) \\ 
\multicolumn{4}{c}{Fit~A} \\
$\rho^+$ & $1.$ (fixed) & $0.$ (fixed) & $76.5 \!\pm\! 1.8 \!\pm\! 2.5$  \\
$\rho^0$ & $0.56\!\pm\! 0.02 \!\pm\! 0.03$ & $10 \!\pm\! 3 \!\pm\! 2$ & $23.9 \!\pm\! 1.8
\!\pm\! 2.1$  \\
$\rho^-$ & $0.65 \!\pm\! 0.03 \!\pm\! 0.02$ & $176 \!\pm\! 3 \!\pm\! 2$ & $32.3 \!\pm\! 2.1
\!\pm\! 1.3$  \\
NR & $1.03 \!\pm\! 0.17 \!\pm\! 0.12$ & $77 \!\pm\! 8 \!\pm\! 5$ & $2.7 \!\pm\!
0.9 \!\pm\! 0.2$  \\ 
& & & $<\!6.4$ @ 95\% C.L.\\ \hline
\multicolumn{4}{c}{Fit~B1}  \\
$\rho^+$ & $1.$ (fixed) & $0.$ (fixed) & $76.6 \!\pm\! 2.5$  \\
$\rho^0$ & $0.57\!\pm\! 0.03$ & $10 \!\pm\! 4$ & $24.9 \!\pm\! 2.4$  \\
$\rho^-$ & $0.64 \!\pm\! 0.03$ & $176 \!\pm\! 4$ & $31.0 \!\pm\! 2.8$  \\
NR & $1.03 \!\pm\! 0.24$ & $72 \!\pm\! 11$ & $2.8 \!\pm\! 1.4$  \\ \hline
\multicolumn{4}{c}{Fit~B2}  \\
$\rho^+$ & $1.$ (fixed) & $0.$ (fixed) & $76.0 \!\pm\! 2.7$  \\
$\rho^0$ & $0.55\!\pm\! 0.04$ & $9 \!\pm\! 4$ & $22.5 \!\pm\! 2.7$  \\
$\rho^-$ & $0.67 \!\pm\! 0.04$ & $177 \!\pm\! 4$ & $34.0 \!\pm\! 3.02$  \\
NR & $1.03 \!\pm\! 0.24$ & $84 \!\pm\! 11$ & $2.7 \!\pm\! 1.4$  \\ \hline
\multicolumn{4}{c}{Fit~C1}  \\
$\rho^+$ & $1.$ (fixed) & $0.$ (fixed) & $78.0 \!\pm\! 2.1$  \\
$\rho^0$ & $0.56\!\pm\! 0.02$ & $9 \!\pm\! 3$ & $24.4 \!\pm\! 1.9$  \\
$\rho^-$ & $0.66 \!\pm\! 0.03$ & $176 \!\pm\! 3$ & $33.9 \!\pm\! 2.3$  \\
$\sigma(500)$ & $0.22 \!\pm\! 0.06$ & $355 \!\pm\! 24$ & $0.08 \!\pm\! 0.08$  \\ 
& & & $<\!0.21$ @ 95\% C.L.\\ \hline
\multicolumn{4}{c}{Fit~C2}  \\
$\rho^+$ & $1.$ (fixed) & $0.$ (fixed) & $78.3 \!\pm\! 1.8$  \\
$\rho^0$ & $0.56\!\pm\! 0.02$ & $10 \!\pm\! 3$ & $24.9 \!\pm\! 1.9$  \\
$\rho^-$ & $0.66 \!\pm\! 0.03$ & $178 \!\pm\! 3$ & $33.4 \!\pm\! 2.1$  \\
$f_0(980)$ & $0.074 \!\pm\! 0.025$ & $325 \!\pm\! 23$ & $0.010 \!\pm\! 0.008$  \\ 
& & & $<\!0.026$ @ 95\% C.L.\\ \hline
\multicolumn{4}{c}{Fit~D}  \\
$\rho^+$ & $1.$ (fixed) & $0.$ (fixed) & $76.3 \!\pm\! 1.9 \!\pm\! 2.5 $  \\
$\rho^0$ & $0.57\!\pm\! 0.03 \!\pm\!0.03$ & $10 \!\pm\! 3 \!\pm\!2 $& $24.4 \!\pm\! 2.0\!\pm\!2.1$  \\
$\rho^-$ & $0.67 \!\pm\! 0.03 \!\pm\! 0.02$ & $178 \!\pm\! 3\!\pm\!2.0$ & $34.5 \!\pm\! 2.4\!\pm\!1.3$  \\
$K$-matrix  & $0.70 \!\pm\! 0.20\!\pm\!0.12$ & $2 \!\pm\! 14\!\pm\!5$ & $0.9 \!\pm\! 0.7\!\pm\!0.2$ \\
& & & $<\!1.9$ @ 95\% C.L.\\ \hline
\end{tabular}
\end{table}

We calculate an integrated $CP$ asymmetry across the Dalitz plot as
described in Ref.~\cite{cpvkspipi} as 
the difference between the integral of the coherent sum of all
amplitudes across the Dalitz plot for $\DZ$ and $\DZB$, respectively, 
divided by the sum of the integral of the coherent sum of all
amplitudes across the Dalitz plot for $\DZ$ and $\DZB$, respectively,
divided by the area of the Dalitz plot.
We obtain ${\cal A}_{CP}\!= 0.01^{+0.09}_{-0.07} \pm 0.05 $,
where the errors are
statistical and systematic, respectively.

We consider systematic uncertainties from experimental sources and
from the decay model separately.
Contributions to the experimental systematic uncertainties arise
from our
model of the background, the efficiency, the signal fraction and the
event selection.
Our general procedure is to change some aspect of Fit~A or Fit~D and
interpret the change in the values of the amplitudes, phases, and fit
fractions as an
estimate of the systematic uncertainty.
In Fit~A and Fit~D, we fix the coefficients of the background determined
from a sideband region. To estimate the systematic uncertainty on 
this background shape we perform a fit with the coefficients allowed
to float constrained by the covariance matrix of the background fit.
We use the covariance matrix of the efficiency fit to estimate the
systematic uncertainty due to the efficiency parameterization. We
generate samples of efficiency parameters based on the covariance
matrix and re-run the Dalitz-plot fit for each sample.
We change selection criteria in the
analysis to test whether our simulation properly models the efficiency.
We vary the minimum $\pi^0$ daughter energy criteria, the cuts on $Q$
and $M$, and the $D^{*+}$ minimum momentum fraction.
We take the square root of the 
sample variance of the amplitudes, phases and fit fractions
from the nominal result compared to the results in
this series of fits as a measure of the
experimental systematic uncertainty. 

We consider the uncertainty
arising from the choice of the $\pi\pi$
$S$-wave model included in the fit.
We interpret the variation in
the $\rho$ amplitudes, phases, and fit fractions in Fit~A, C1, C2 and
D as a modeling systematic uncertainty. We add the experimental and model
systematic uncertainty in quadrature to obtain the total systematic
uncertainty reported in Table~\ref{tab:fit}.

	In conclusion, we have analyzed the resonant substructure of 
the decay $\DZ \to \pi^+\pi^-\pi^0$
using the Dalitz-plot analysis technique.
We observe the three $\rho(770)\pi$ resonant and a small non-resonant
contribution. We find no evidence for a $\pi\pi$ $S$-wave contribution
with either the Breit-Wigner or $K$-matrix parameterization. 
We find $A_{CP}$ which is the asymmetry between the $D^0$
and $\overline{D^0}$ distributions integrated over the entire Dalitz
plot to be $0.01^{+0.09}_{-0.07} \pm 0.05 $.  


We gratefully acknowledge the effort of the CESR staff 
in providing us with
excellent luminosity and running conditions.
This work was supported by 
the National Science Foundation and
the U.S. Department of Energy.

\end{document}